\documentclass[twocolumn,pra,amsmath]{revtex4}

\usepackage{graphicx}

\def\onlinecite{\cite}

\begin{document}

\title{Low-energy dynamical response of an Anderson insulator with local attraction}

\author{D.~A.~Ivanov$^{1,2}$ and M.~V.~Feigel'man$^{3,4}$}

\affiliation{$^{1}$ Institute for Theoretical Physics, ETH Z\"urich, 8093 Z\"urich, Switzerland}
\affiliation{$^{2}$ Department of Physics, University of Z\"urich, 8057 Z\"urich, Switzerland}
\affiliation{$^{3}$ L.~D.~Landau Institute for Theoretical Physics, Chernogolovka, 142432,
Moscow region, Russia}
\affiliation{$^{4}$  National Research University ``Higher School of Economics'', Moscow, Russia}

\begin{abstract}
The low-frequency dynamical response of an Anderson insulator is dominated
by so-called Mott resonances: hybridization of pairs of states close in energy, but
separated spatially. We study the effect of interaction on Mott resonances in
the model of spinful fermions (electrons) with local attraction. This model is known
to exhibit a so-called pseudogap: a suppression of the low-energy single-particle excitations.
Correspondingly, the low-energy dynamical response is also reduced. However this
reduction has mostly quantitative character. In particular, the Mott formula
for frequency-dependent conductivity preserves its functional asymptotic
behavior at low frequencies, but with a small numerical prefactor. This result
can be explained in terms of Mott resonances for electron pairs instead of single
electrons.
\end{abstract}

\date{December 17, 2016}

\maketitle

\section{Introduction}

Adding interactions to an Anderson insulator \cite{anderson-1958} brings in a broad spectrum of new
phenomena, from Coulomb gap and related repulsion effects \cite{efros-1975,shklovskii-1981,efros-1985}
to many-body localization \cite{basko-2006,MBLreview}.
Recently, a model of a disordered metal with local {\it attraction} was studied in 
Refs.\ \onlinecite{feigelman-2007,feigelman-2010}.
This model exhibits an superconductor-insulator transition (SIT)~\cite{IM2010,FIM2010}, 
with unusual properties of the superconducting state formed by paired localized electrons.
The insulating state on the other side of the SIT is expected to be unusual as well; in
particular, it may realize a very good testground for experimental studies of
many-body localization phenomena~\cite{FIM2010}.

The latter development motivated us to study in more detail the effect of local attraction on the
low-energy dynamic response in an Anderson insulator. It is known that low-energy properties of
a coherent Anderson insulator are determined by Mott resonances: hybridization of pairs of states
close in energy, but separated spatially \cite{mott-1970}. In particular, the dynamical density-response
function can be calculated using the statistical distribution of Mott resonances \cite{ivanov-2012}, and,
in 1D, these calculations reproduce the results obtained directly with the Gorkov-Berezinskii method
\cite{gorkov-1983}. In higher dimensions. the statistical properties of Mott resonances are less studied,
and similar results can only be conjectured \cite{ivanov-2012}. This dynamical density-response function
can be used to express physically relevant quantities such as frequency-dependent conductivity and
polarizability.

In this paper, we study the same dynamical density-response function in the presence of local attraction,
similar to the model considered in Refs.~\onlinecite{feigelman-2007,feigelman-2010}.
The interaction introduces an additional
energy scale: the pseudogap energy $\Delta_P$. One therefore expects (and it will be confirmed by the calculation
below) that the dynamical density-response function is affected by interaction at frequencies below $\Delta_P$
(we put $\hbar=1$ throughout the paper and do not distinguish between energy and frequency). 
We consider the regime $\Delta_P \ll \Delta_\xi$, where $\Delta_\xi$
is the level spacing at the localization length $\xi$ (i.e., the energy scale associated with localization).
We also consider the zero-temperature limit (temperature much lower than all other energy scales).
Finally, we assume the absence of superconductivity (the insulating side
of the superconductor-insulator transition~\cite{feigelman-2007,feigelman-2010}).

Under all these assumptions, we study the dynamical density-response function in the whole range of frequencies
$\omega \ll \Delta_\xi$ (without any assumption on the relation between $\omega$ and $\Delta_P$). As expected,
at $\omega \gg \Delta_P$, the dynamical density-response function is close to its non-interacting form, but
at $\omega \ll \Delta_P$ it is considerably reduced. This reduction can be viewed as two renormalizations:
first, the overall numerical prefactor is renormalized; second, the Mott length scale is renormalized (approximately
by the factor of one half). This can be interpreted in terms of Mott resonances for pairs of particles.

The paper is organized as follows. 
In Section \ref{sec:chi}, we define the main object of our study: the dynamical density-response function
and discuss its general properties. In Section \ref{sec:model}, we define the model of the disordered
system with interaction and its simplified version studied in the paper. In Section \ref{sec:mott}, we
derive a general formula for the contribution to the dynamical density-response function from Mott
resonances. Section \ref{sec:cases} solves two special cases of such resonances, the non-interacting case
and the perfect-resonance case, which are further used as a basis for a more general analysis. In
Section \ref{sec:low-energy}, we compute the effect of the interaction at frequencies $\omega$ much lower than
the pseudogap energy $\Delta_P$. Section \ref{sec:numeric} contains numerical results on the renormalization
of the density-response function in the more general situation of $\omega$ comparable to $\Delta_P$.
Finally, in Section \ref{sec:discussion}, we discuss conditions of applicability of our results
and their physical implications for the frequency-dependent conductivity and polarizability.

\section{Dynamical density-response function}
\label{sec:chi}

We define the disorder-averaged dynamical density-response function at zero temperature 
(more precisely, its dissipative component) as
\begin{equation}
s(\omega, x) = \frac{1}{\omega} \left\langle
\sum_i \left\langle 0 | {\hat n}(0) | i \right\rangle 
\left\langle i | {\hat n}(x) | 0 \right\rangle
\delta(E_i-E_0-\omega) \right\rangle\, ,
\label{linear-response-1}
\end{equation}
where
\begin{equation}
{\hat n}(x) = \sum_\sigma \hat\Psi^\dagger_\sigma(x) \hat\Psi_\sigma(x)
\end{equation}
is the fermionic density operator ($\hat\Psi^\dagger_\sigma(x)$ and $\hat\Psi_\sigma(x)$ are
the fermionic creation and annihilation operators, $\sigma$ is the spin index), $\left| 0 \right\rangle$
and $\left| i \right\rangle$ are the ground state and the excited eigenstates with the energies
$E_0$ and $E_i$, respectively, $\langle\ldots\rangle$ denotes disorder averaging.

This response function is related to various physical observables.
For example, the frequency-dependent conductivity (the dissipative part) can
be expressed as
\begin{equation}
\sigma_{\alpha\beta} (\omega) = - \frac{\pi}{2} \omega^2 \int dx \; x_\alpha x_\beta \; s(\omega, x)
\label{sigma-def}
\end{equation}
and the static polarizability, via the Kramers-Kronig relation, as 
\begin{equation}
\chi_{\alpha\beta} = 
\frac{2}{\pi} \int_0^\infty \frac{d\omega}{\omega^2} \; \sigma_{\alpha\beta}(\omega) =
- \int dx \int d\omega  \; x_\alpha x_\beta \; s(\omega, x)
\label{chi-def}
\end{equation}
(here and below the integral over $x$ is understood in the $d$-dimensional space).

The dimensionality of the so defined response function $s(\omega, x)$
is (Energy$\cdot$Volume)$^{-2}$. For non-interacting fermions, in the approximation
of a constant density of states,
\begin{equation}
s(\omega, x)_{\rm nonint}= 2\nu^2 S(\omega,x)\, ,
\label{relation-to-noninteracting}
\end{equation}
where $\nu$ is the density of states {\em per spin} and $S(\omega,x)$ is the single-particle
response function as defined in Ref.~\onlinecite{ivanov-2012} (for spinless fermions):
\begin{multline}
  S(\omega, x)
  =
  \nu^{-2}\,
  \Big\langle
  \sum_{n,m} \delta(E_n-E) \delta(E_m-E-\omega)
\\
  \psi^*_n(0) \psi_n(x) \psi^*_m(x) \psi_m(0)
  \Big\rangle \, ,
\label{S-function}
\end{multline}
where $\psi_n(x)$ are single-particle eigenstates.
The coefficient 2 in Eq.~(\ref{relation-to-noninteracting})
comes from the spin degeneracy.

\begin{figure}[tb]
\centerline{\includegraphics[width=0.3\textwidth]{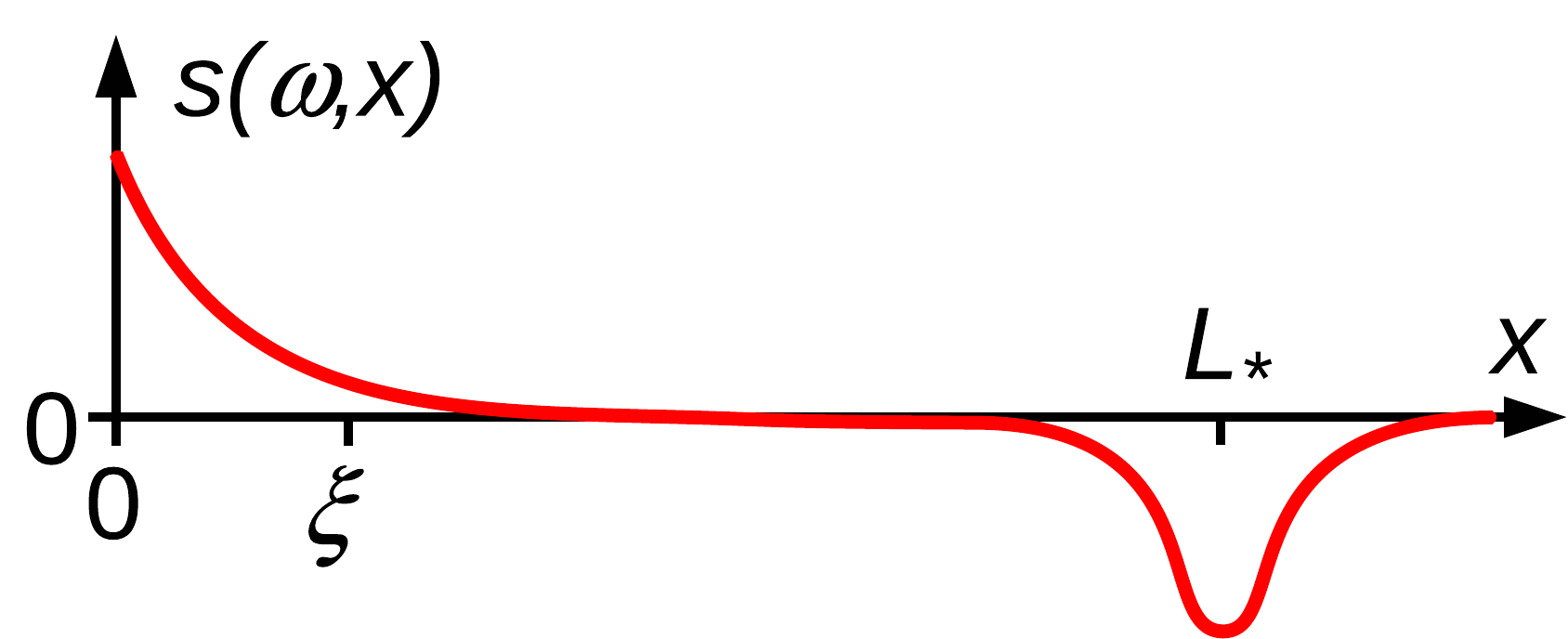}}
\caption{A schematic view of the dynamical density-response function $s(\omega,x)$.
The position of the hump $L_*$ is the Mott length scale (\ref{L-mott}) in
the non-interacting case, but is renormalized by interaction.}
\label{fig:s-plot-general}
\end{figure}

The qualitative shape of the dynamical density-response function $s(\omega, x)$ at $\omega \ll \Delta_\xi$ 
is shown in Fig.~\ref{fig:s-plot-general}. At small $x$ ($x\lesssim\xi$),  
$s(\omega, x)$ is proportional to the correlation function of a {\em single} localized state
\cite{ivanov-2012}. At large $x$ ($x\gg \xi$), $s(\omega, x)$ has a negative hump, so that the overall
integral obeys the sum rule (following from the orthogonality of eigenstates):
\begin{equation}
\int dx\; s(\omega, x)=0\, .
\label{sum-rule}
\end{equation}
For non-interacting fermions in one dimension, this form of $s(\omega, x)$ was derived in 
Ref.~\onlinecite{gorkov-1983}. Later, it was reproduced in Ref.~\onlinecite{ivanov-2012} based
on the Mott-hybridization phenomenology, where it was further conjectured that a similar form should also hold
in higher dimensions.

In the non-interacting case, the $x$ position of the negative hump is given by the Mott length
\cite{mott-1970,gorkov-1983,ivanov-2012}
\begin{equation}
L_M=2\xi\ln(\Delta_\xi/\omega)\, ,
\label{L-mott}
\end{equation}
and this is the only dependence of $s(\omega,x)$ on $\omega$ in the small-$\omega$ limit.

We argue below that this form of $s(\omega,x)$ remains valid in a model with local attraction,
with the two differences: the $x$ position and the overall weight are renormalized.

\section{Model}
\label{sec:model}

We adopt the model of Refs.~\onlinecite{feigelman-2007,feigelman-2010}, which includes
electrons with disorder and with local attraction between electrons of
opposite spin. 
Attraction between electrons of the same spin has little effect
on the considered phenomena and can be neglected.
One possible form of the system Hamiltonian is
\begin{equation}
\hat{H}=\hat{H}_{\rm kin}+\hat{H}_{\rm disorder}+\hat{H}_{\rm int}\, ,
\end{equation}
where the kinetic, disorder, and interaction parts of the Hamiltonian are
\begin{equation}
\hat{H}_{\rm kin} = \int dx\; 
\sum_\sigma \hat\Psi^\dagger_\sigma(x) \left(-\frac{\nabla^2}{2m}\right) \hat\Psi_\sigma(x) \, ,
\end{equation}
\begin{equation}
\hat{H}_{\rm disorder} = \int dx\; 
\sum_\sigma \hat\Psi^\dagger_\sigma(x) V_{\rm disorder}(x) \hat\Psi_\sigma(x) \, ,
\end{equation}
\begin{equation}
\hat{H}_{\rm int} =  - g \int dx\; 
\hat\Psi^\dagger_\uparrow(x) \hat\Psi^\dagger_\downarrow(x) \hat\Psi_\downarrow(x) \hat\Psi_\uparrow(x) \, .
\end{equation}
The specific form of the kinetic term, the statistics of the disorder potential, and of the
short-range attraction is not important for the discussion below, since only the long-distance
properties of the tails of the localized states are relevant. In particular, it will be convenient
to consider the model in the basis of localized states,
\begin{multline}
\hat{H} =
\sum_i \varepsilon_i \left(\hat\Psi^\dagger_{i\uparrow} \hat\Psi_{i\uparrow} + 
\hat\Psi^\dagger_{i\downarrow} \hat\Psi_{i\downarrow}\right) \\
+ \sum_{i\ne j} J_{ij} \left(\hat\Psi^\dagger_{i\uparrow} \hat\Psi_{j\uparrow} + 
\hat\Psi^\dagger_{i\downarrow} \hat\Psi_{j\downarrow}\right) \\
- \sum_i \Delta_{i} \left(\hat\Psi^\dagger_{i\uparrow} \hat\Psi^\dagger_{i\downarrow}
\hat\Psi_{i\downarrow} \hat\Psi_{i\uparrow}\right)\, .
\label{mott-resonance-H}
\end{multline}
Here the operators $\hat\Psi_{i\sigma}$ correspond to states localized at the localization
length $\xi$, without including long-range Mott hybridization. The small tunneling amplitudes
$J_{ij}$ describe long-range Mott hybridization, and the attraction is included as the one-site
energy term $\Delta_i$ favoring double occupation of localized states. We
neglect attraction between different localized states, assuming the system to be 
far from the SIT. Furthermore,
to simplify the calculation, we first assume that the
energy gain from a double occupation $\Delta_i$ is the same
for all localized states (and denote it $\Delta_P$).
In reality, this energy gain is different for
different states and depends on the inverse participation ratio for
a given state \cite{feigelman-2010}. We discuss the effect of the distribution
of $\Delta_i$ qualitatively after the main calculation.

\begin{figure}[tb]
\centerline{\includegraphics[width=0.3\textwidth]{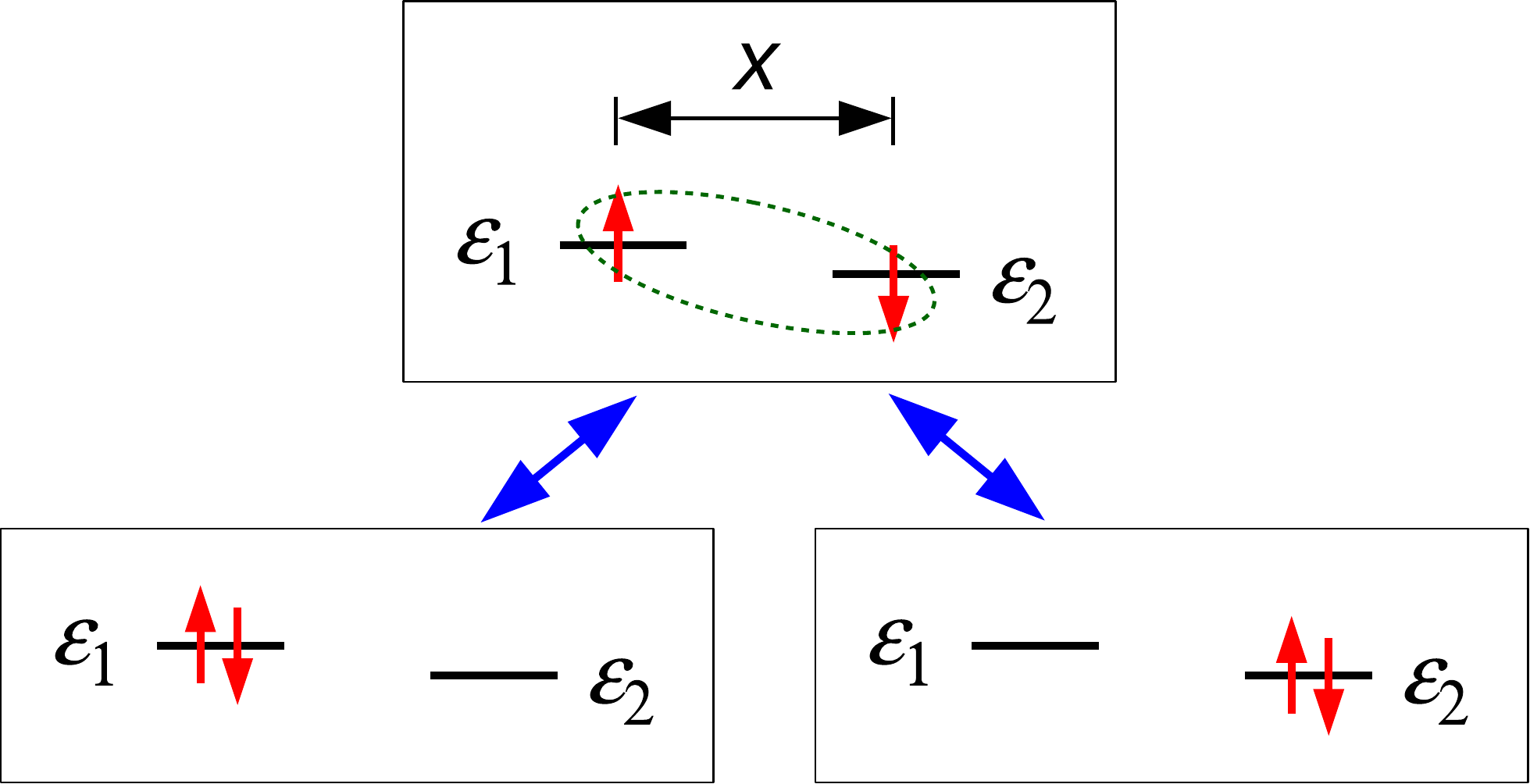}}
\caption{The three states used in the Mott-resonance calculation.
The green dashed ellipse denotes the spin-singlet antisymmetrization.
The blue double-ended arrows indicate possible transition between the states.}
\label{fig:three-states}
\end{figure}

\section{Low-energy contribution from Mott-resonant states}
\label{sec:mott}

Even in the presence of interactions, provided $\omega \ll \Delta_\xi$, the 
dynamical density-response function $s(\omega,x)$
at low frequencies $\omega$ is determined by Mott resonances. Below we calculate it 
by solving the case of a single Mott resonance 
 [the Hamiltonian (\ref{mott-resonance-H}) with two states only] 
and then averaging over the
statistics of such Mott resonances, following the method of Ref.~\onlinecite{ivanov-2012}.
A similar approach was used in Refs.\ \onlinecite{shklovskii-1981,efros-1985}
to treat the case of Coulomb repulsion. Unlike that calculation where spin played
no important role, the case of attraction requires an explicit treatment of the spin
degrees of freedom.

Consider a Mott resonance formed by two localized states at initial energies $\varepsilon_1+\Delta_P/2$ 
and $\varepsilon_2+\Delta_P/2$ located at the positions $0$ and $x$, respectively (we
include the energy shift of $\Delta_P/2$ for convenience). 
We assume $x\gg \xi$. Let the Mott-hybridization
amplitude equal $J$. This hybridization amplitude is a random variable
(depending on the disorder realization and on the particular pair of states), whose
probability distribution may, in principle, depend on the type of disorder.
It was conjectured in Ref.\ \onlinecite{ivanov-2012} that, in a single-parameter-scaling
regime, the probability distribution is approximately log-normal:
\begin{equation}
dP_x(J) \approx \sqrt{\frac{\xi}{\pi x}}
\exp \left[ - \frac{(2\xi \ln|J/\Delta_\xi|+x)^2}{4 x\xi} \right]\; d\ln |J|\, .
\end{equation}
For our derivation below, the actual form of the probability distribution $dP_x(J)$
is not important: the only thing we assume about it is that its width in the logarithmic
scale is much larger than unity (we believe that it is a plausible assumption at distances
$x\gg \xi$ for a wide class of disorder models). We further
assume $\omega \ll \Delta_\xi$.

The two localized states produce four single-particle states (due to the spin degeneracy),
which can be occupied in the ground state by a number of particles between 0 and 4.
One can show, however, that, in the case of attractive interaction, the ground state 
must always have an even number of particles
(0, 2, or 4). 
Indeed, if a one-particle state has a negative energy, one can lower
the energy even further by adding one more particle in the same state, but with an opposite
spin; the same argument applies for a three-particle state, if one uses holes instead of particles.
The state with 0 particles has energy 0, the state with 4 particles has
energy $2(\varepsilon_1+\varepsilon_2)$. If one of those states has energy lower
than the lowest two-particle state, this resonance does not contribute, since there are no
relevant excitations.

The only contribution to $s(\omega, x)$ comes from the resonances with the
two-particle ground state. Because of the particle-hole symmetry, we can assume 
$\varepsilon_1+\varepsilon_2>0$ and then multiply the result by $2$ (to include the
sector with $\varepsilon_1+\varepsilon_2<0$). Furthermore, it is easy to see that
the relevant two-particle states are in the singlet sector (the triplet sector has
the energy $\varepsilon_1+\varepsilon_2+\Delta_P$, which is higher than that of at least one
of the states with 0 particles or with 4 particles). There are three such states
with the normalized basis
\begin{equation}
\frac{1}{\sqrt2}
(\hat\Psi^\dagger_{1\uparrow} \hat\Psi^\dagger_{2\downarrow}
-\hat\Psi^\dagger_{1\downarrow} \hat\Psi^\dagger_{2\uparrow}) 
\left| \star \right\rangle\, , \quad
\hat\Psi^\dagger_{1\uparrow} \hat\Psi^\dagger_{1\downarrow} \left| \star \right\rangle \, , \quad
\hat\Psi^\dagger_{2\uparrow} \hat\Psi^\dagger_{2\downarrow} \left| \star \right\rangle\, , 
\end{equation}
where $\left| \star \right\rangle$ is the vacuum state (without particles), 
and $\hat\Psi^\dagger_{\alpha\sigma}$ are the electron creation operators in the localized state $\alpha$
with the spin $\sigma$, see Fig.~\ref{fig:three-states}.
The Hamiltonian restricted to these three states has the form
\begin{equation}
H=\varepsilon_1+\varepsilon_2 + H_0\, \qquad
H_0=\begin{pmatrix}
\Delta_P & \sqrt2 J & \sqrt2 J \\
\sqrt2 J & \varepsilon & 0 \\
\sqrt2 J & 0 & -\varepsilon \\
\end{pmatrix}\, ,
\label{H0-matrix}
\end{equation}
where $\varepsilon=\varepsilon_1-\varepsilon_2$. The contribution of such states
to the linear-response function (\ref{linear-response-1}) can be written as
\begin{multline}
s(\omega, x) = \frac{2 \nu^2}{\omega} 
\int dP_x(J)
\iint_{\varepsilon_1+\varepsilon_2>0}
d\varepsilon_1 \; d\varepsilon_2 
\\
\theta(-(\varepsilon_1+\varepsilon_2+E_0))\; 
\sum_{i=1,2} \left\langle 0 |n_1 | i \right\rangle \left\langle i |n_2 | 0 \right\rangle
\delta(E_i-E_0-\omega)\, .
\end{multline}
Here, $\nu$ is the density of states at the Fermi level {\em per spin},
$n_1={\rm diag}(1,2,0)$ and $n_2={\rm diag}(1,0,2)$ are the number-of-particles operators
in the first and the second localized state, respectively. The state $\left| 0 \right\rangle$
is the lowest of the three eigenstates of the Hamiltonian $H_0$ (with the eigenvalue $E_0$),
and the states $\left| 1 \right\rangle$ and $\left| 2 \right\rangle$ are the two
excited states.

Integration over $\varepsilon_1+\varepsilon_2$ may be easily performed. It is further
convenient to replace the operators $n_1$ and $n_2$ by the operator $D=(n_1-n_2)/2=
{\rm diag(0,1,-1)}$, using the orthogonality of the eigenstates. We arrive at
\begin{multline}
s(\omega, x) =-2 \nu^2  \int dP_x(J)
\int_0^\infty d\varepsilon 
\\
\frac{(-E_0)}{\omega}
\sum_{i=1,2} | \left\langle 0 | D | i \right\rangle |^2 \; \delta(E_i-E_0-\omega)
\label{linear-response-3}
\end{multline}
(here we have folded the integral, without loss of generality, on the half line $\varepsilon>0$).

\section{Special cases}
\label{sec:cases}

\subsection{Non-interacting case}

We first check that the above formula reproduces the non-interacting result in the limit $\Delta_P=0$.
In this case, the Hamiltonian $H_0$ can be diagonalized with the eigenvalues
\begin{equation}
E_0=-\sqrt{\varepsilon^2 + 4J^2} \, , \quad
E_1=0 \, , \quad
E_2=\sqrt{\varepsilon^2 + 4J^2}\, . 
\end{equation}
On explicitly computing the eigenvectors and substituting them into (\ref{linear-response-3})
(note that, in the non-interacting case, $\left\langle 0 | D | 2 \right\rangle=0$, which simplifies
the calculation),
we find
\begin{equation}
s(\omega, x)_{\rm nonint}= -\nu^2 \int_{2|J|<\omega} dP_x(J) \frac{4J^2}{\omega\sqrt{\omega^2-4J^2}}\, ,
\label{s-nonint-1}
\end{equation}
which reproduces Eq.~(31) of Ref.~\onlinecite{ivanov-2012} (up to a factor of 2 due to spin).
The limit of integration in $|J|$ in Eq.~(\ref{s-nonint-1}) emerges naturally as a consequence
of the delta-function constraint in Eq.~(\ref{linear-response-3}).

For completeness, we reproduce further calculation from Ref.~\onlinecite{ivanov-2012}.
Note that the integration measure $dP_x(J)$ is distributed on the logarithmic scale of $J$
with a width much larger than one, while the integrand has the width of order unity on the
same logarithmic scale. We may therefore
replace the integral (\ref{s-nonint-1}) by 
\begin{equation}
s(\omega, x)_{\rm nonint}= -C\nu^2 \int dP_x(J) \; \delta(\ln|J| - \ln\omega)\, ,
\label{s-nonint-2}
\end{equation}
where the coefficient $C$ is calculated as
\begin{equation}
C=\int_{2|J|<\omega} d\ln|J| \frac{4J^2}{\omega\sqrt{\omega^2-4J^2}}=1 \, .
\end{equation}

\subsection{Perfect resonance: $\varepsilon=0$}

A helpful insight in the interaction effect on Mott resonances can be
obtained from the case $\varepsilon=0$ (perfect resonance). In this
case, the Hamiltonian $H_0$ can also be easily diagonalized with the
eigenvalues
\begin{equation}
E_0=-\frac{\Delta_J-\Delta_P}{2}\, , \quad
E_1=0\, , \quad
E_2= \frac{\Delta_J+\Delta_P}{2}\, ,
\label{resonant-eigenvalues}
\end{equation}
where
\begin{equation}
\Delta_J= \sqrt{\Delta_P^2 + 16 J^2}\, .
\end{equation}
Note that in the limit $J\ll \Delta_P$, the excitation gap is small: 
$E_0 \approx -4J^2/\Delta_P$, in spite of the single-particle gap.

\section{Low-energy limit: $\omega\ll \Delta_P$}
\label{sec:low-energy}

The dynamical density-response function (\ref{linear-response-3}) 
may be calculated analytically in the limit $\omega\ll \Delta_P$. In this
case,
as will be checked self-consistently, the relevant scale of $|J|$ also
obeys $|J|\ll \Delta_P$. In this regime,
the first of the three basis states in the matrix (\ref{H0-matrix}) is separated by a large energy
gap from the other two states and can be taken into account perturbatively. The effective
Hamiltonian for the remaining two states is
\begin{equation}
H_{\rm eff} = \begin{pmatrix}
\varepsilon-J_{\rm eff} & -J_{\rm eff} \\
-J_{\rm eff} & -\varepsilon-J_{\rm eff}
\end{pmatrix}\, , \qquad
J_{\rm eff} = 2J^2/\Delta_P\, ,
\label{H-eff}
\end{equation}
and its diagonalization gives the energies
\begin{equation}
E_0=-J_{\rm eff}-\sqrt{\varepsilon^2+J_{\rm eff}^2}\, , \quad
E_1=-J_{\rm eff}+\sqrt{\varepsilon^2+J_{\rm eff}^2}\, .
\end{equation}
Continuing the calculation along the lines of the non-interacting case considered above
[specifically, using Eq.~(\ref{linear-response-3}) with only one excited state
and substituting the eigenvectors of the 2$\times$2 Hamiltonian (\ref{H-eff})],
we find
\begin{equation}
s(\omega, x) = -C(\omega) \nu^2 \int dP_x(J) \;
\delta(\ln|J| - \ln J_*(\omega))\, ,
\end{equation}
where the characteristic overlap scale $J_*(\omega)$ is now
\begin{equation}
J_*(\omega\ll \Delta_J) = \sqrt{\omega\Delta_P}
\label{J-low-omega}
\end{equation}
(which, indeed, obeys $J_*\ll\Delta_P$),
and the total weight is
\begin{multline}
C(\omega) =2 \int_{2J_{\rm eff}<\omega} d\ln|J|\; \frac{J_{\rm eff}^2}{\omega^2}\;
\sqrt{\frac{\omega+2J_{\rm eff}}{\omega-2J_{\rm eff}}} \\
= \frac{4+\pi}{16} \approx 0.446\ \, .
\end{multline}

This result is very similar to the noninteracting case, except for two differences. 
First, the overall numerical coefficient is renormalized. Second, the Mott length scale (\ref{L-mott})
is replaced by $L_*=2\xi \ln(\Delta_\xi/\sqrt{\omega\Delta_P})$ in the low-frequency limit of
the interacting case.

\begin{figure}[tb]
\centerline{\includegraphics[width=0.47\textwidth]{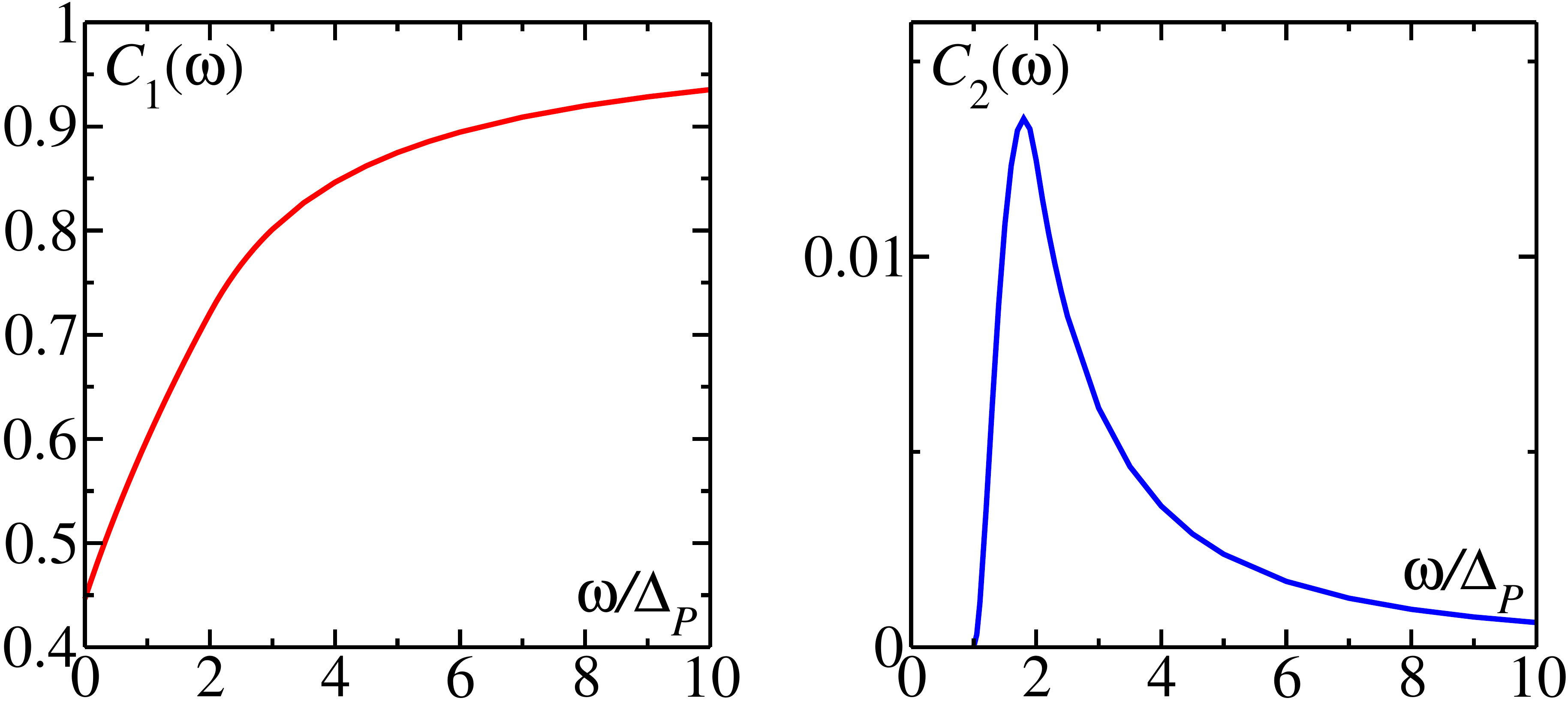}}
\caption{Numerically calculated coefficients $C_1(\omega)$ and $C_2(\omega)$.}
\label{fig:C1-C2}
\end{figure}

\section{General case}
\label{sec:numeric}

In the general case ($\omega$ comparable to $\Delta_P$, but much smaller than $\Delta_\xi$), the
expression (\ref{linear-response-3}) cannot be calculated analytically, and we need to resort to
a numerical study. In addition, for a numerically accurate estimate one would need to take into
account the statistical distribution of $\Delta_P$, which was neglected in earlier formulas.
The main part of the
calculation will be performed at a fixed $\Delta_P$, and the averaging over $\Delta_P$ will
only be discussed qualitatively afterwards.

As in the previous sections, one can argue that, since $J$ is distributed on a logarithmic scale,
one can write in general
\begin{equation}
s(\omega, x) = s_1(\omega, x) +s_2(\omega, x)\, ,
\end{equation}
where $s_1(\omega, x)$ and $s_2(\omega, x)$ are the contributions from the two
excited states in Eq.~(\ref{linear-response-3}):
\begin{equation}
s_i(\omega, x)= -C_i(\omega) \nu^2 \int dP_x(J)\; \delta(\ln|J| - \ln J^{(i)}_*(\omega))\, ,
\end{equation}
where $J^{(i)}_*(\omega)$ are the characteristic scales of the relevant tunnelling amplitudes.
With logarithmic precision, they may be estimated from the zero-$\varepsilon$ case
by solving $E_i-E_0=\omega$, where $E_i$ and $E_0$ are given by 
Eq.~(\ref{resonant-eigenvalues}). For the main excitation ($i=1$), this gives
(up to a coefficient $1/2$ which is beyond our logarithmic level of precision)
\begin{equation}
J^{(1)}_*(\omega) \sim \sqrt{\omega(\omega+\Delta_P)}\, .
\end{equation}
Alternatively, this formula may be understood as an interpolation between the
low-frequency regime (\ref{J-low-omega}) and the high-frequency noninteracting
limit $J_*=\omega$ [see eq.~(\ref{s-nonint-2})]. The second contribution ($i=2$)
has $J^{(2)}_*(\omega) \sim \sqrt{\omega^2-\Delta_P^2}$, which is of the same order
of magnitude as $J^{(1)}_*(\omega)$, except in a narrow vicinity of $\Delta_P$
(the second contribution only appears for $\omega>\Delta_P$).

The renormalization coefficients $C_i(\omega)$ are given by
\begin{multline}
C_i(\omega)=2\int d\ln|J| \int_0^\infty d\varepsilon \\
 \frac{(-E_0)}{\omega}
| \left\langle 0 | D | i \right\rangle |^2 \; \delta(E_i-E_0-\omega)\, .
\label{C-i-general}
\end{multline}
Results of the numerical calculation of $C_1(\omega)$ and $C_2(\omega)$ are shown
in Fig.~\ref{fig:C1-C2}. First, even though $C_2(\omega)$ has a strict
cut-off at $\omega=\Delta_P$, its actual contribution needs to be averaged over
a distribution of $\Delta_P$, so that no sharp feature is expected in 
$s(\omega, x)$ as a function of $\omega$. Second, the
numerical values of $C_2(\omega)$ are very small, as compared to $C_1(\omega)$, 
so for most practical purposes the contribution $s_2(\omega, x)$ can be neglected. 
Third, the renormalization coefficient $C_1(\omega)$ further needs
to be averaged over the distribution of $\Delta_P$, which would produce a function
qualitatively similar to the left panel of Fig.~\ref{fig:C1-C2}, but
quantitatively slightly smoothed (the $\omega \to 0$ limit remains the same).

\section{Physical implications}
\label{sec:discussion}

\subsection{Dynamical response function}

Our results can be summarized in a simple relation between the dynamical response
function with and without interaction:
\begin{equation}
s(\omega \ll \Delta_\xi, x) = C(\omega/\Delta_P)\; s(\sqrt{\omega(\omega+\Delta_P)}, x)_{\rm nonint}\, ,
\label{interacting-noninteracting}
\end{equation}
where $C(\omega/\Delta_P)$ is a function increasing from $0.446$ to $1$ on the characteristic
scale of $\omega \sim \Delta_P$. Note that even though our derivation above only considers
the region $x\gg \xi$, the relation (\ref{interacting-noninteracting}) is valid for all $x$, due
to the sum rule (\ref{sum-rule}) and to the fact the the correlation function of a single localized
state is only weakly modified by interaction as long as $\Delta_P\ll\Delta_\xi$.

The exact form of the function $C(\omega/\Delta_P)$ should depend on the number
of dimensions, since it involves averaging $C_1(\omega)$ over 
the statistics of $\Delta_P$. The latter, in turn, depends on the statistics
of the inverse participation ratio, which is dimension dependent; 
such a statistics in various dimensions was studied in
Refs.~\onlinecite{fyodorov-1994,fyodorov-1995,mirlin-2000,feigelman-2010}.
The results reported there suggest that the width of the distribution of $\Delta_P$
is of the order of $\Delta_P$ itself, but in two and three dimensions it can only be obtained
from numerical simulations. Such an averaging would require numerical methods beyond
the simple analytical approach of the present paper. As a result, it would only smooth 
$C_1(\omega)$ without changing its qualitative behavior (in particular, preserving the
$\omega\to 0$ limit). For this reason, we do not perform it here, but only discuss qualitatively. 

The function  $s(\omega, x)_{\rm nonint}$ is known analytically in 1D \cite{gorkov-1983}, 
but is only conjectured in higher dimensions \cite{ivanov-2012}.

\subsection{Frequency-dependent conductivity}

The result (\ref{interacting-noninteracting}) implies a similar relation for the frequency-dependent
conductivity (\ref{sigma-def}):
\begin{equation}
\sigma(\omega \ll \Delta_\xi) = C(\omega/\Delta_P)\; \frac{\omega}{\omega+\Delta_P}
\sigma(\sqrt{\omega(\omega+\Delta_P)})_{\rm nonint}\, .
\label{interacting-noninteracting-conductivity}
\end{equation}
The well-known Mott argument based on the structure of $s(\omega, x)$ shown in Fig.~\ref{fig:s-plot-general} 
predicts the low-frequency behavior \cite{mott-1970}
\begin{equation}
\sigma(\omega) \propto \omega^2(\ln \omega)^{d+1}
\label{sigma}
\end{equation}
in $d$ dimensions. Our result (\ref{interacting-noninteracting-conductivity}) implies that, 
while this form is preserved in the presence of interactions at $\omega \ll \Delta_P$,
the overall proportionality coefficient is strongly reduced by the factor $0.446 \cdot (1/2)^{d+1}$, which
is about $0.03$ in the three-dimensional case.

\subsection{Polarizability}

From the above discussion, one can also estimate an effect of interaction on the  static polarizability $\chi$.
In the presence of interaction, the integrand in Eq.~(\ref{chi-def}) is strongly reduced below the energy
scale $\omega\sim\Delta_P$. As a consequence, the polarizability is reduced as compared to the noninteracting
case. The magnitude of this interaction correction, in the regime $\Delta_P \ll \Delta_\xi$, can be
estimated, using Eqs.\ (\ref{chi-def}) and (\ref{sigma}), as
\begin{equation}
\frac{\delta\chi}{\chi} \propto - \frac{\Delta_P}{\Delta_\xi} \left( \ln \frac{\Delta_\xi}{\Delta_P}\right)^{d+1}\, ,
\end{equation}
with a numerical proportionality coefficient of order one. The calculation presented above is insufficient
to calculate this proportionality coefficient, as it would require, in particular, the knowledge of
the statistical distribution of the interaction energies $\Delta_i$.

\subsection{Universality of results}

The main result (\ref{interacting-noninteracting}) has a high level of universality:
the only assumption used in its derivation was that the probability distribution of the
hybridization matrix element $J$ is a slow function of $\ln |J|$. We have not assumed any
specific details of this probability distribution. Therefore, we believe that our result
is applicable to any type of disorder and in any dimension. However, the exact form of the
function $C(\omega/\Delta_P)$ may depend on the details of the disorder statistics
and on the dimensionality.

In our derivation, we assumed zero temperature. This assumption was crucial for excluding single-particle
hopping and keeping only hopping of pairs (Fig.~\ref{fig:three-states}). Thus the validity of
 our results is restricted by the condition $T \ll \Delta_P$.  Even in this range, some additional $T$-dependent
contribution to conductivity due to thermally excited pair-tunnelling processes can be expected, 
similar to the Austin-Mott law~\cite{AustinMott}; we leave this subject for a future study.

\subsection{Possible experiments}

In terms of measurable quantities, our main result (\ref{interacting-noninteracting-conductivity})
provides a prediction for the frequency dependence of the {\em ac} conductivity in the
insulating state of materials that possess local attraction between electrons. 
By now, the best-studied example of this type is the amorphous non-stoichiometric 
indium oxide. It is well known that InO$_x$ with conduction electron densities
of the order of a few $10^{20}$ cm$^{-3}$ is superconducting
or insulating depending upon a slight variation of the oxygen content or upon applying a magnetic field
of the order of several Teslas \cite{exp1,exp2,Sacepe2015}. A pseudogap in this material was demonstrated,
in particular, by scanning tunnelling spectroscopy in the superconducting state with a typical
transition temperature around 1-2 K \cite{Sacepe2011}.  A relatively weak magnetic field is known to
suppress such a fragile superconductivity, leading to an insulating state with a giant resistivity 
above $10^9$ Ohm at temperatures below 0.1 K \cite{exp2,Tamir}. An even stronger insulating state is
known to appear in the same InO$_x$ compound with somewhat higher resistivity \cite{exp1}; 
a purely activated resistivity proportional to $\exp(T_0/T)$ was reported in Ref.~\onlinecite{Kowal94}
with $T_0$ as high as 15 K. These examples demonstrate that measurements of real and imaginary parts of the
dielectric response in the GHz range and at sub-Kelvin temperatures should be able to experimentally test
our theoretical predictions.

\acknowledgments

We thank E.~Cuevas and B.~Shklovskii for discussions.
The work of D.A.I.\ was supported by the
Swiss National Foundation through the NCCR QSIT.
The research was also partially supported by the RF Presidential Grant 
No.\ NSh-10129.2016.2.

\end{document}